\documentclass[aps,prl,preprintnumbers,twocolumn,superscriptaddress,nofootinbib]{revtex4-1}
\usepackage{amsmath,amsthm,amssymb,color,psfrag,url,latexsym,graphicx,epstopdf,slashed,xspace,hyperref,enumitem}
\usepackage{lineno}

\definecolor{darkred}{rgb}{0.6,0.0,0.0}
\definecolor{darkblue}{rgb}{0.0,0.0,0.5}
\definecolor{darkgreen}{rgb}{0.0,0.5,0.0}
\definecolor{brown}{rgb}{0.0,0.0,0.0}

\newcommand{\be}{\begin{equation}}
\newcommand{\ee}{\end{equation}}
\newcommand{\bea}{\begin{eqnarray}}
\newcommand{\eea}{\end{eqnarray}}

\begin{document}
\preprint{YITP-SB-2021-17}
\title{Probing hadronization with flavor correlations of leading particles in jets}

\author{Yang-Ting Chien}
\email{yang-ting.chien@stonybrook.edu}
\email{ytchien@gsu.edu}
\affiliation{Center for Frontiers in Nuclear Science, Stony Brook University, Stony Brook, NY 11794}
\affiliation{Department of Physics and Astronomy, Stony Brook University, Stony Brook, NY 11794}
\affiliation{C.N. Yang Institute for Theoretical Physics, Stony Brook University, Stony Brook, NY 11794}
\affiliation{Physics and Astronomy Department, Georgia State University, Atlanta, GA 30303}

\author{Abhay Deshpande}
\email{abhay.deshpande@stonybrook.edu}
\affiliation{Center for Frontiers in Nuclear Science, Stony Brook University, Stony Brook, NY 11794}
\affiliation{Department of Physics and Astronomy, Stony Brook University, Stony Brook, NY 11794}
\affiliation{Brookhaven National Laboratory, Upton, NY 11973}

\author{Mriganka Mouli Mondal}
\email{mrigankamouli.mondal@stonybrook.edu}
\affiliation{Center for Frontiers in Nuclear Science, Stony Brook University, Stony Brook, NY 11794}
\affiliation{Department of Physics and Astronomy, Stony Brook University, Stony Brook, NY 11794}

\author{George Sterman}
\email{george.sterman@stonybrook.edu}
\affiliation{Center for Frontiers in Nuclear Science, Stony Brook University, Stony Brook, NY 11794}
\affiliation{Department of Physics and Astronomy, Stony Brook University, Stony Brook, NY 11794}
\affiliation{C.N. Yang Institute for Theoretical Physics, Stony Brook University, Stony Brook, NY 11794}

\date{\today}

\begin{abstract}
We study nonperturbative flavor correlations between pairs of leading and next-to-leading charged hadrons within jets at the Electron-Ion Collider (EIC). We introduce a charge correlation ratio observable $r_c$ that distinguishes same- and opposite-sign charged pairs. Using Monte Carlo simulations with different event generators, $r_c$ is examined as a function of various kinematic variables for different combinations of hadron species, and the feasibility of such measurements at the EIC is demonstrated. The precision hadronization study we propose will provide new tests of hadronization models and hopefully lead to improved quantitative, and perhaps eventually analytic, understanding of nonperturbative QCD dynamics.
\end{abstract}
\maketitle

The analysis of high energy collisions requires the treatment of both perturbative and non-perturbative aspects of the strong interactions.   Many observable processes, especially those involving jet production and energy flow substructure are fundamentally perturbative \cite{Larkoski:2017jix}.   The universality of non-perturbative jet evolution reflected in single-particle inclusive data requires a description that combines perturbative coefficients with non-perturbative fragmentation functions \cite{Metz:2016swz}. At the other extreme, the prediction of fully exclusive final states, involving multiple identified particles, requires the full technology of event generators including models of hadronization \cite{Field:1977fa, Andersson:1997xwk, PhysRevLett.59.1997, Webber:1983if, Winter:2003tt}. Accurate and systematically improvable perturbative calculations and parton shower Monte Carlo (MC) simulations \cite{Giele:2007di, Nagy:2007ty, Schumann:2007mg, Dasgupta:2020fwr} are also essential for describing partonic distributions before hadronization\footnote{While lattice quantum chromodynamics (QCD) methods have made great strides in calculating hadronic properties such as mass spectra, QCD phase diagram, parton distribution functions (PDFs) and other relevant QCD matrix elements \cite{Kronfeld:2012uk}, the application of these methods to hadronization remains for the future.}. In this note, we point out that the tagging of particle flavor and momenta within jets can provide robust sets of observables that are dependent on the dynamics of hadronization. Such observables can test existing models and perhaps lead to new insights.    

In order to extract localized hadronization features, we focus in this study exclusively on the leading and next-to-leading energy hadrons within jets and examine their origin from a boosted, intrinsically nonperturbative system surrounding high energy partons \footnote{In contrast, the abundant soft particles may collectively come from all possible soft event activities, whose specific perturbative or nonperturbative origins are challenging to identify individually \cite{Stewart:2014nna, Chien:2019osu}.}.  Given the identified leading hadron $H_1$ and next-to-leading hadron $H_2$, the two-particle correlation spans a kinematic phase space \cite{Chen:2019apv}, encoding the conditional probability of observing $H_2$ in the presence of $H_1$. We define a charge correlation ratio, $r_c$, from the differential cross sections ${\rm d}\sigma_{H_1H_2}/{\rm d}X$ to quantify flavor and kinematic dependence of hadronization in the production of $H_1=h_1$ and $H_2=h_2$ or $\overline{h_2}$ (the anti-particle of $h_2$),
\begin{equation}
	r_c (X)= \frac{{\rm d}\sigma_{h_1h_2}/{\rm d}X-{\rm d}\sigma_{h_1\overline{h_2}}/{\rm d}X}{{\rm d}\sigma_{h_1h_2}/{\rm d}X+{\rm d}\sigma_{h_1\overline{h_2}}/{\rm d}X}\;.
	\label{eq:r_c-def}
\end{equation}
We will explore the dependence of $r_c$ on a variety of kinematic variables, $X$. In the defintion, Eq.\ (\ref{eq:r_c-def}), $H_1$ and $H_2$ can in principle be arbitrary hadron species, including charged and neutral hadrons. 

We first focus on the correlations among charged hadrons, since they can be identified using efficient tracking \cite{AbdulKhalek:2021gbh}, with significant progress in the phenomenology and in theoretical frameworks for track-based observables \cite{Krohn:2012fg, Waalewijn:2012sv, Chang:2013rca, Chang:2013iba, Chien:2020hzh, Kang:2020fka, Kang:2021ryr, Li:2021zcf}. In this case, the ratio $r_c$ is constructed with the convention that the electric charges of $h_1$ and $h_2$ have the same sign. Note that the ratio satisfies $-1 \le r_c \le 1$, and that it approaches $-1$ when ${\rm d}\sigma_{h_1h_2}/{\rm d}X \ll {\rm d}\sigma_{h_1\overline{h_2}}/{\rm d}X$, i.e., when the cross section with opposite-sign leading dihadrons is dominant. The ratio is expected to be negative on combinatoric grounds, because of charge conservation. This is typically the case, although the effects we find are much larger than simple combinatorics would suggest.    A similar charge density ratio was introduced by the TASSO experiment long ago \cite{TASSO:1985hiu}, as well as in the studies of Refs.\ \cite{CERN-CollegedeFrance-Heidelberg-Karlsruhe:1978krl, ACCDHW:1979mjc, EHSNA22:1989cad, TASSO:1981gag, TASSO:1982bkc, TPCTwoGamma:1984bra, TPCTwoGamma:1984tkq, TPCTwoGamma:1986yqu, OPAL:1993wav}, for inclusive particle correlations, typically in relative rapidity. Also, a closely-related ``balance function" was employed in heavy ion studies \cite{Bass:2000az, ALICE:2013vrb} to investigate the different hadronization environment in those experiments. In contrast to these studies, we model high statistics experiments at the EIC, to compute our ratio, $r_c$ specifically for flavor- and energy-tagged hadrons in a variety of kinematic variables in their relative momenta.

As the collision energy increases, jets emerge from enhanced emissions of soft and collinear partons. While many jet properties at short distance can be accurately described by perturbative tools with moderate hadronization corrections, we focus on identifying specific jet substructures that are predominantly determined by hadronization. With a fundamental mismatch between their quantum numbers, multiple partons and hadrons are necessarily involved during hadronization, even in the simplest kinematic configuration. The Pythia Lund string model \cite{Ferreres-Sole:2018vgo} and the Herwig cluster model \cite{Bahr:2008pv} are two successful examples of hadronization prescriptions implemented in MC simulations. Each requires pre-hadronization partons to connect with strings or form individual clusters, which then turn into multiple hadrons. Certainly, the large partonic phase space spanned in a high energy collision will result in a complicated string or cluster configuration, which will be reflected in the details of the hadronic final states, including jets and soft particles. The leading dihadrons are a special subset of jet particles, which can illuminate some intrinsic dynamics of hadronization.

On the theory side, hadronic momentum distributions can be organized systematically using generalized multi-hadron fragmentation functions \cite{Majumder:2004wh,Majumder:2004br, Elder:2017bkd}. We can extend the fragmentation formalism to leading and next-leading hadrons in jets \cite{Procura:2009vm, Jain:2011iu, Kaufmann:2015hma, Chien:2015ctp, Kang:2016ehg, Dai:2016hzf, Elder:2017bkd}.  Such distributions would receive contributions from the perturbative emission of partons, and from their subsequent fragmentation. Relative to single-particle distributions, however, the distributions of leading dihadrons have enhanced sensitivity to their nonperturbative origin.  For example, hadronization through string dynamics conserves electric charge locally, i.e., string breaking creates an electric charge neutral quark and anti-quark pair.   This pattern implies strong constraints on the flavor and charge content of the dihadrons, with a preference for them to carry oppositely charged valence quarks.

\begin{figure}[b]
    \centering
    \includegraphics[trim=.2in 0.0in  0.1in 0.0in, width=0.45\textwidth]{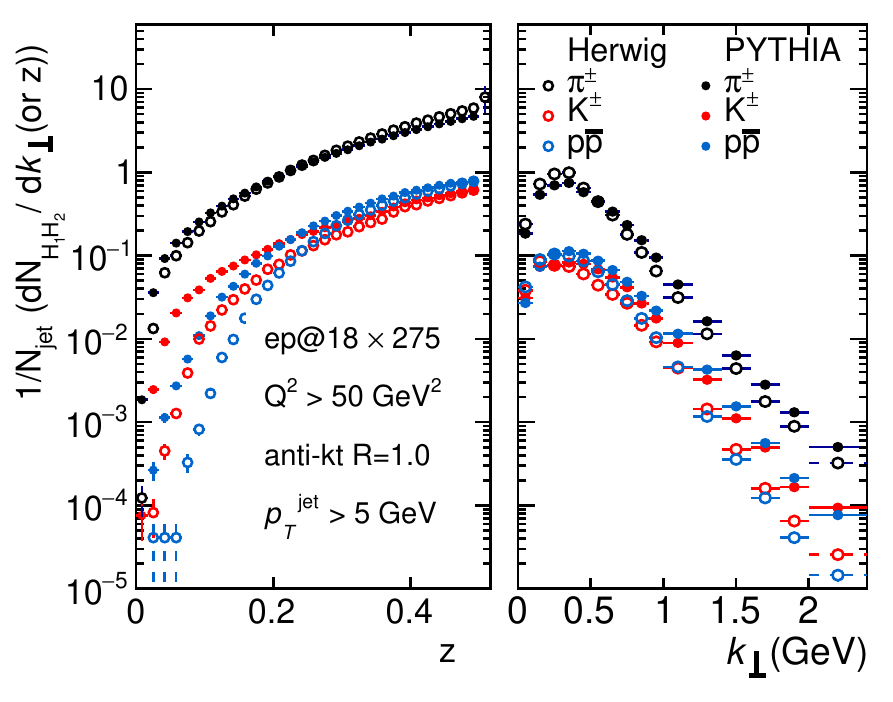}
    \caption{Leading dihadron $z$ (left panel) and $k_\perp$ (right panel) distributions for $\pi^\pm$ (black), $K^\pm$ (red) and $p\bar p$ (blue) from PYTHIA (solid circles) and Herwig (open circles) simulated jets..}
    \label{fig:kine}
\end{figure}

\begin{figure*}[t]
    \centering
    \includegraphics[trim=.1in .1in 0.1in 0.1in, width=0.99\textwidth]{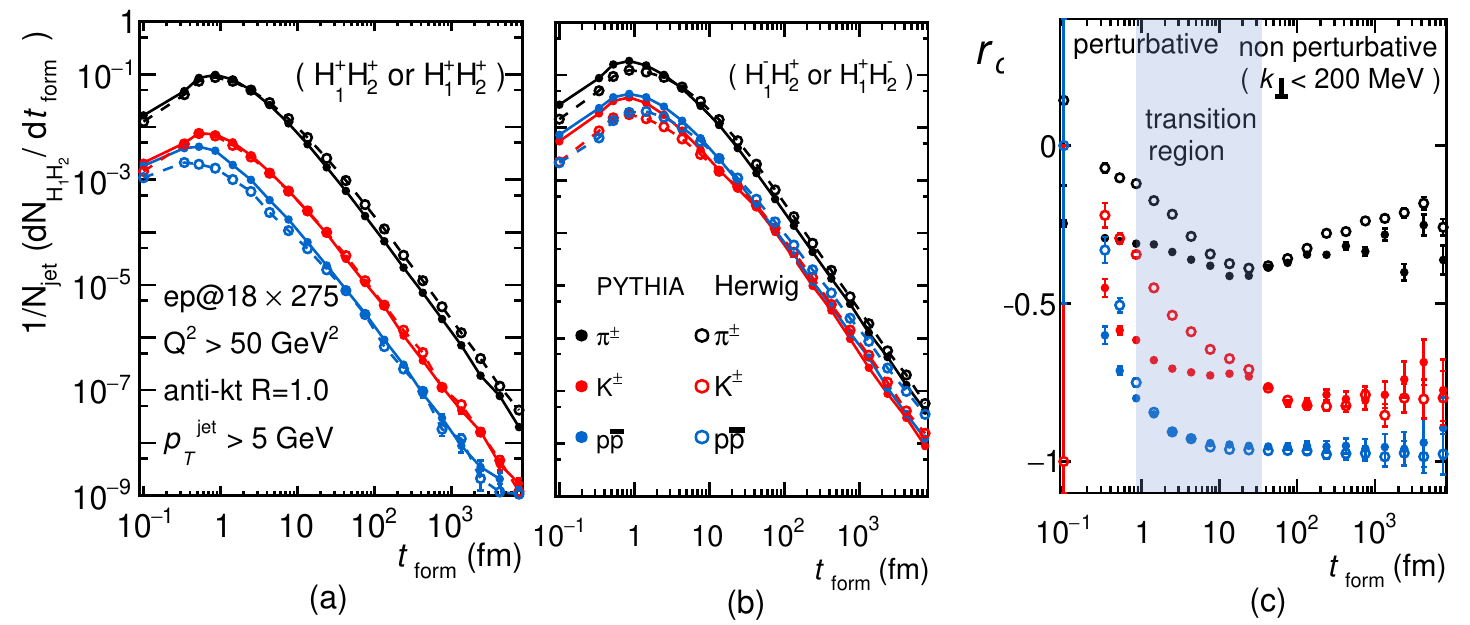}
    \caption{Leading dihadron formation time $t_{\rm form}$ distributions for $\pi^\pm$ (black), $K^\pm$ (red) and $p\bar p$ (blue) from PYTHIA (solid circles) and Herwig (open circles) simulated jets. Panel (a) shows the distributions for $H_1^+H_2^+$ or $H_1^-H_2^-$ with the same-sign electric charge, while panel (b) shows the ones for $H_1^-H_2^+$ or $H_1^+H_2^-$ with opposite-sign electric charges. The distributions for $K^\pm$ and $p\bar p$ are scaled up by a factor of 3. Panel (c) shows the charge correlation $r_{c}$ as a function of $t_{\rm form}$.}
    \label{fig:tform}
\end{figure*}

We focus on kinematic regions which include both relatively low- and high-energy jets. The next generation Electron Ion Collider (EIC), with polarized beams and tunable beam energies, will provide us an ideal laboratory for such studies. It is planned to operate and start taking data by 2030, with collider and detector designs currently under active discussion.  We study the benchmark highest achievable EIC energies of 18 GeV electron beam and 275 GeV proton beam. The studies are performed based on Monte Carlo simulations of Deep Inelastic Scattering (DIS) events using PYTHIA 6.428 \cite{Sjostrand:2006za} and Herwig 7.1.5 \cite{Bellm:2015jjp} event generators. An event selection with DIS kinematics $Q^2 > 50~{\rm GeV}^2$ is imposed so that jets have moderately high transverse momenta, and we analyze 10 million such events, which corresponds to approximately one percent of the expected integrated luminosity at the EIC.

Jets are reconstructed using the anti-$k_t$ algorithm \cite{Cacciari:2008gp} with $R=1.0$ using \textsc{FastJet} 3~\cite{Cacciari:2011ma}. We include particles with transverse momenta above 0.2 GeV and pseudo rapidity within the range $-1.5  \le \eta \leq  3.5$. We consider jets with transverse momenta $p_T^{\rm jet} > 5$ GeV and with leading and next-to-leading particles both charged. We assume that these leading particles can be identified and will discuss detector requirements. In realistic measurements at the EIC, one needs to consider detector acceptance, tracking efficiency and momentum resolution factors. The EIC fast simulation package \cite{AbdulKhalek:2021gbh} has been developed for simulating the quality of a measurement. For the flavor correlation studies, one needs high momentum resolution for the two leading tracks, with precise charge determination and particle identification of charged pions and kaons. The proton collision data at RHIC (STAR \cite{STAR:2005gfr} and future sPHENIX \cite{PHENIX:2015siv}) and the LHC \cite{ALICE:2008ngc, CMS:2008xjf, ATLAS:2008xda}, as well as archived DIS data at HERA (H1 \cite{H1:1996jzy}) and $e^{+}e{-}$ collision data at LEP \cite{DELPHI:1990cdc, ALEPH:1990ndp, Badea:2019vey, Chen:2021iyj} are also available. However, they are lack of precise $\pi/K$ separation at high momentum, which makes the EIC unique for such measurements.

We will examine the dependence of $r_c$ within the relative kinematic phase space of the leading and next-leading hadrons.   The formation time \cite{Dokshitzer:1991wu} $t_{\rm form} = z(1-z)p /k_\perp^2$ observed in the laboratory frame gives information about the spacetime picture of leading dihadron production. Here, the $p= p^{H_1}+p^{H_2}$ is the total leading dihadron momentum, $z=p^{H_2}/p$ is the longitudinal momentum fraction of the softer hadron, and $k_\perp$ is the relative transverse momentum between the two leading particles. We sum over initial and final states with different polarizations, and leave the study of hadronization for polarized states for future work. 

FIG. \ref{fig:kine} shows the inclusive distributions in $z$ and $k_\perp$. They are clearly nonperturbative in origin, with the curve in $z$ rising sharply away from $z=0$, and $k_\perp$ falling exponentially.  If the two leading hadrons originate from an intrinsically nonperturbative process in their rest frame, their relative transverse momentum should be at a nonperturbative energy scale while the formation time would be Lorentz dilated. Also, the Lorentz boost effect implies that the two hadrons may tend to carry comparable momenta so that $z$ is of the order of $1/2$, while their relative $k_\perp$ would remain at a nonperturbative scale. In contrast, $z\approx 1/2$ is disfavored if the next-to-leading hadron comes from the fragmentation of perturbative, soft emission.

FIG. \ref{fig:tform} shows on a logarithmic scale the distributions of the leading dihadron formation time $t_{\rm form}$ for $\pi^\pm$, $K^\pm$ and $p\bar p$, with their electric charges of the same sign ($H_1^+H_2^+$ or $H_1^-H_2^-$, panel (a)) or opposite signs ($H_1^-H_2^+$ or $H_1^+H_2^-$, panel (b)). The $K^\pm$ and $p\bar p$ distributions are scaled up by a factor of 3 for readability on the same plot. In comparing the two left panels, we can already see the dominance of opposite-charged pairs over same charges.

The distributions in FIG. \ref{fig:tform} peak between 1 and 10 fm and decrease at large ($\gtrsim10$ fm) and small ($\lesssim 1$ fm) formation time.  Panel (c) shows the charge correlation ratio $r_c$ as a function of $t_{\rm form}$. With the sign convention, we see that $r_c$ is mostly negative, with significant differences in the correlations among the hadron species. It is highly unlikely to produce same sign $pp$ or $\bar p\bar p$ compared to $p\bar p$. Also, it is much more likely to observe two leading kaons with opposite signs due to strangeness conservation in the production of $s\bar s$ quark pair. If an energetic sea $s$ (or $\bar s$) quark is struck out of the proton\footnote{The other $\bar s$ (or $s$) goes along the beam direction and escapes jet reconstruction.}, its flavor correlation with another $s\bar s $ pair which contributes to producing the other energetic kaon within the jet should be weak perturbatively. The strong $K^\pm$ correlation indicates either strong nonperturbative flavor constraints, or that the two leading kaons are produced by a single $s\bar s$ pair from gluon splitting or string breaking. There is a weaker tendency to produce opposite-sign leading pions. For large formation time where the dynamics is dominated by non-perturbative physics, the strength of the correlation is stronger and flat. This region typically corresponds to energetic, collinear dihadrons with $k_\perp \lesssim 200$ MeV. On the other hand, correlations are weaker at small formation time, hinting at early time de-correlations for wide-angle, perturbative emissions. The intermediate transition region is then sensitive to the variation of charge correlation strength. Note that Herwig and PYTHIA show distinct features for pions and kaons at $t_{\rm form} \lesssim 10$ fm.

\begin{figure}[t]
    \centering
    \includegraphics[trim=.2in 0.0in  0.1in 0.0in, 
    width=0.48\textwidth]{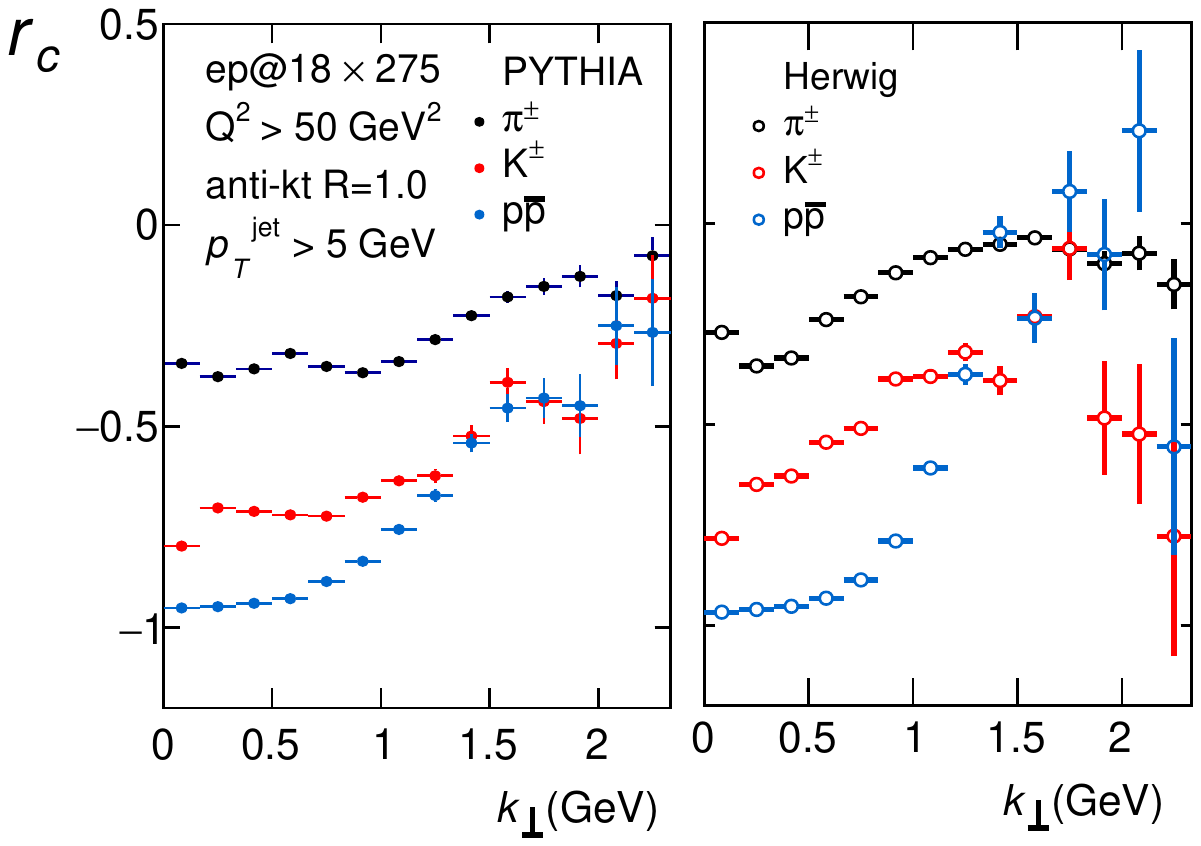} 
    \caption{Charge correlation ratio $r_{c}$ for identified leading dihadrons $\pi^{\pm}$ (black), $K^{\pm}$ (red) and $p\bar{p} $ (blue) as a function of relative $k_\perp$  in PYTHIA (left panel) and Herwig (right panel) simulated jets.}
    \label{fig:kperp}
\end{figure}

The correlations we discuss here are clearly nonperturbative in origin, although in general charge correlations are not always nonperturbative.  At large enough $k_\perp$, perturbative charge correlations between leading dihadrons would depend on universal fragmentation  within different jets or subjets.   For such dihadrons, charge correlations can be inherited from the partons that initiate the jets or subjets, analogously to the case of spin correlations \cite{Webber:1986mc, Collins:1987cp, Knowles:1987cu, Chen:2020adz, Karlberg:2021kwr}, but we anticipate these correlations to be much smaller than the ones we observe here.   This reflects the previous observation that $r_c$ decreases in size as $t_{\rm form}$ vanishes (FIG.\ref{fig:tform}). 

We also examine $r_c$ as a function of $k_\perp$, as shown in FIG. \ref{fig:kperp}. The correlation decreases in absolute value as $k_\perp$ increases on the scale of 1-2 GeV. The description of $r_c$ over this $k_\perp$ range will require both perturbative and nonperturbative inputs.  Detailed comparisons of data and event generator output will help clarify the degrees of freedom necessary to provide a full picture of hadronization throughout this region.

\begin{figure}[t]
    \centering
    \includegraphics[trim=.2in 0.0in  0.2in 0.1in, width=0.48\textwidth]{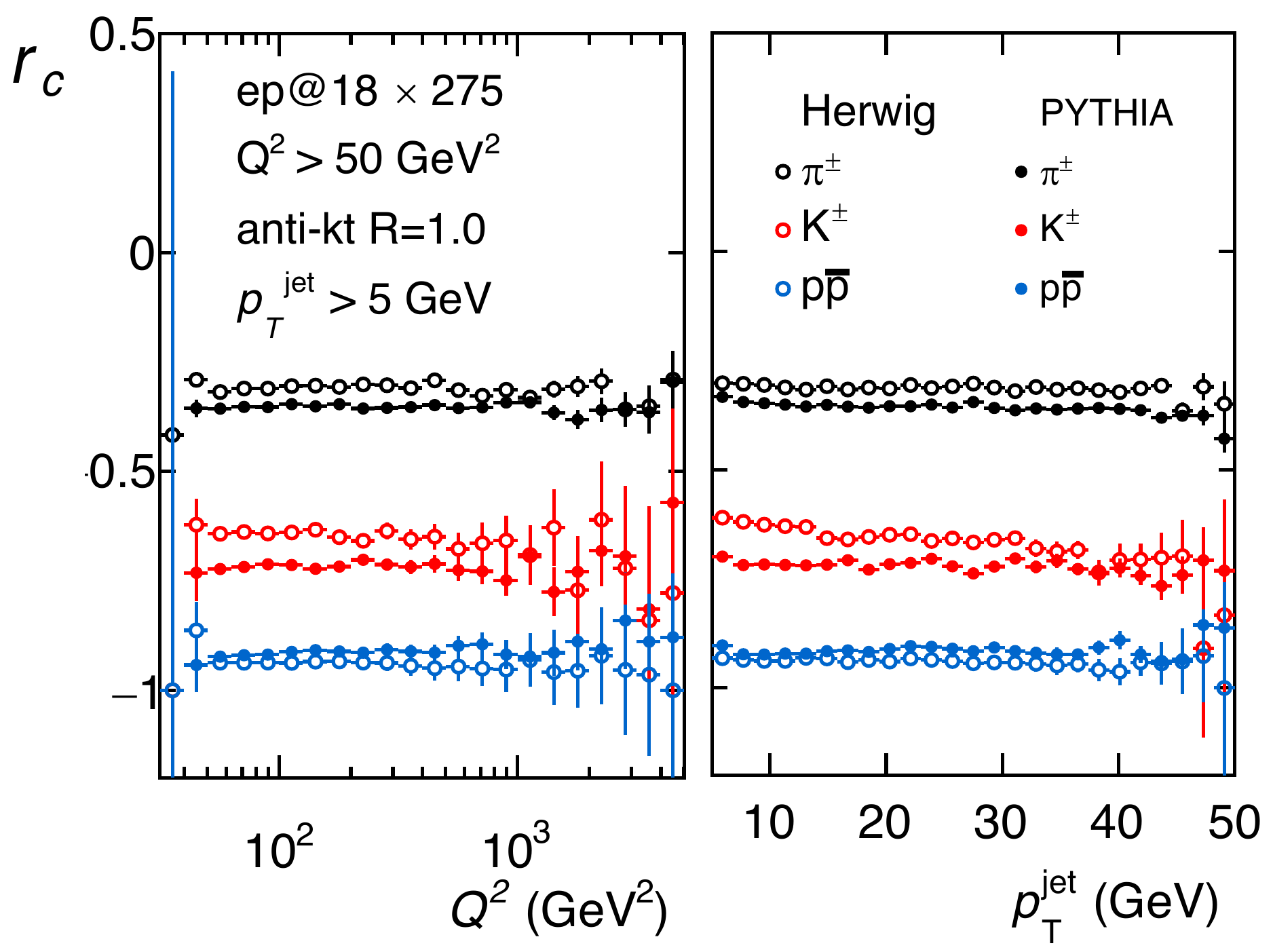}
    \caption{Charge correlation ratio $r_{c}$ for identified leading dihadrons $\pi^{\pm}$ (black), $K^{\pm}$ (red) and $p\bar{p} $ (blue) as a function of hard scales, the DIS $Q^2$ (left panel) and jet $p_T^{\rm jet}$ (right panel) in PYTHIA (solid circle) and Herwig (open circle) simulated jets.}
    \label{fig:Q2pt}
\end{figure}

We show in FIG. \ref{fig:Q2pt} the charge correlation ratio $r_c$ versus two hard scales of the process, $Q^2$ (left panel) and $p_T^{\rm jet}$ (right panel).  Each shows an extraordinary scaling of $r_c$ with these variables.   This behavior of the event generators may reflect a built-in boost invariance of the hadonization process.  In data, whether from the EIC or previous DIS experiments, we might expect a more noticeable evolution with $Q^2$ or $p_T^{\rm jet}$. This is an appealing example, where the high statistics of EIC experiments may provide new tests of the hadronization models built into event generators, and their interface with perturbative showers.

\begin{figure*}[t]
    \centering
    \includegraphics[trim=.0in 0in .0in 0.in, width=0.9\textwidth]{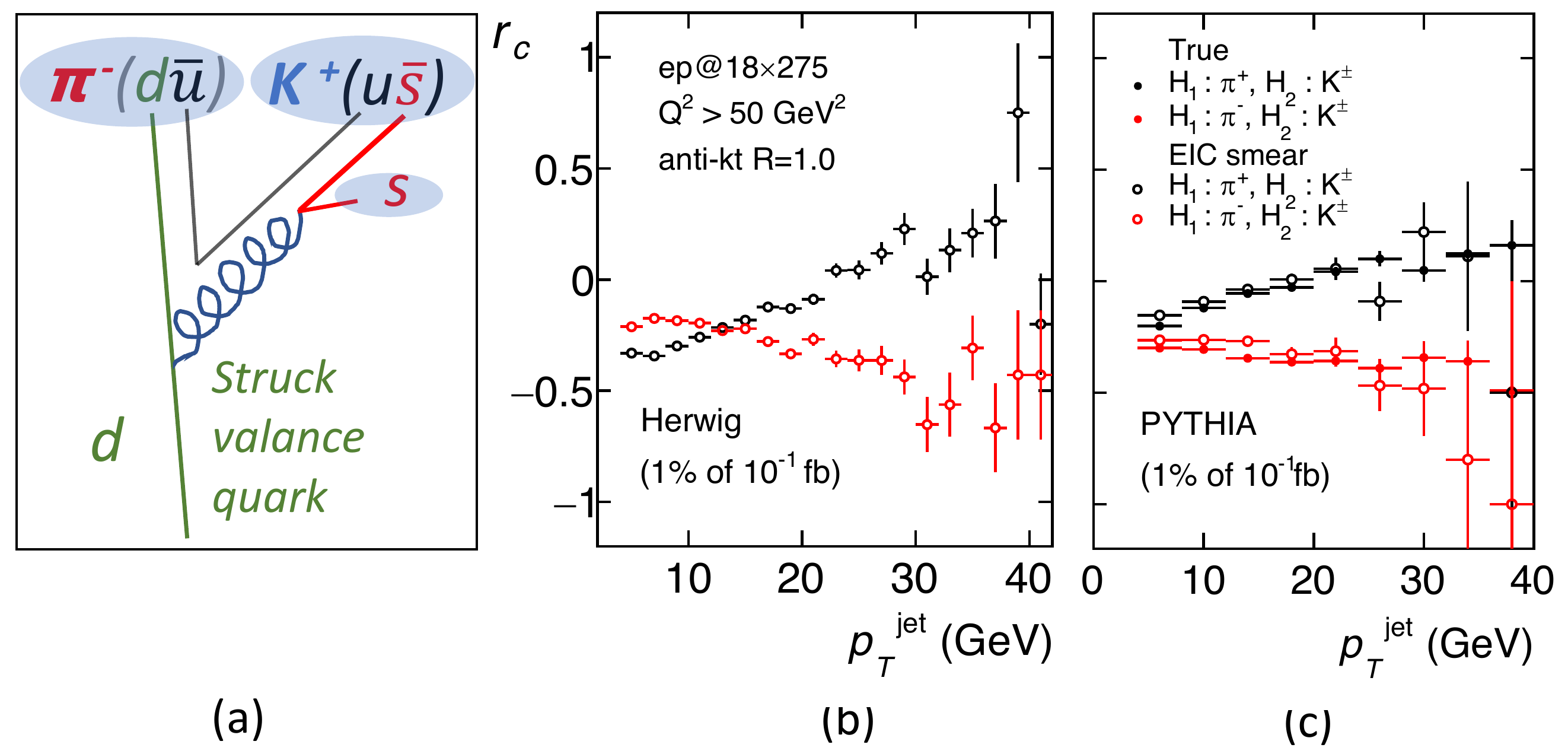}
    \caption{ Panel (a): Illustration of the dominant partonic channel for producing a leading $\pi^-$ and a next-to-leading $K^+$. The struck valance $d$ quark (green) emits a gluon which splits into an $s\bar s$ pair. The string connecting $d$ and $\bar s$ breaks and creates a $\bar u u$ pair which then forms $\pi^-K^+$. Panel (b): Charge correlation ratio $r_c$ as a function of $p_T^{\rm jet}$ for leading $\pi^+K^\pm$ (black) and $\pi^-K^\pm$ (red) in Herwig simulations. Panel (c): Charge correlation ratio $r_c$ as a function of $p_T^{\rm jet}$ for leading $\pi^+K^\pm$ (black) and $\pi^-K^\pm$ (red) in PYTHIA simulations (solid circles) and with EIC detector simulations (open circles). }
    \label{fig:pik}
\end{figure*}

In the case that the struck quark of the DIS process is a valence $u$ or $d$, producing a leading pion, we can study the implications of flavor in hadronization concretely, by requiring a leading, mixed-flavor $\pi K$ correlation. Here, we use string model inspired reasoning. Panel (a) in FIG. \ref{fig:pik} illustrates the dominant partonic channel for producing a leading $\pi^-(d\bar u)$ and a next-to-leading $K^+ (u \bar s)$. A gluon splits to $s\bar s$, either perturbatively or nonperturbatively, and a single string can be formed connecting the struck valence $d$ quark and the $\bar s$ quark. The string breaking can result in the production of $\pi^-K^+$ \footnote{In comparison, a struck $u$ quark and a $\bar d d$ pair from string breaking will form $\pi^-K^0$ which is not included in our current discussion.}. Other hadron species combinations of $\pi^-K^-$ and $\pi^+K^\pm$ need more complicated string configurations and breaking which are phase space and energy disfavored. Therefore $r_c$ is expected to be stronger for $\pi^-K^\pm$ compared to $\pi^+K^\pm$. This expectation is borne out in panel (c) in FIG. \ref{fig:pik}, where the charge correlation indeed is stronger in $\pi^-K^\pm$ compared to $\pi^+K^\pm$ in PYTHIA simulations. On the other hand,  Herwig does not show a similar hierarchy. In the low $p_T^{\rm jet}$ region the correlation strength of $\pi^+K^\pm$ is even stronger than $\pi^-K^\pm$ for Herwig, which is an opposite trend compared to PYTHIA. 

Viewing this qualitative discrepancy, we examine the capability of the EIC with realistic detector simulations, to see if future experimental uncertainty will allow us to distinguish the $\pi K$ correlations. As shown in Panel (c), the PYTHIA EIC smear results are in excellent agreement with the true distributions. The critical part of particle identification at the EIC is at midrapidity where the goal is set to identify $\pi/K$  with 3$\sigma$ separation up to 10 GeV in momentum. Depending on the proton beam energies at the EIC, high-energy jets may be most common at forward rapidity, which may promote $\pi/K$ separation at  relatively high momentum. The EIC is expected to meet such a goal, and possible detector development and R\&D is discussed in the yellow report \cite{AbdulKhalek:2021gbh}. Future measurements at the EIC will thus be able to provide experimental constraints on the hadronization models valid for the $\pi K$ and other correlations.

We close with a few comments on how the studies here can be extended.  Adapting these analyses to archived data from past experiments should be possible, and may already lead to new insights, for example in testing the tight scaling of MC output in $Q^2$ and jet $p_T^{\rm jet}$ (see FIG.\ \ref{fig:Q2pt}).   

Generalizations of $r_c$ to observables that include multiple subleading particles and perhaps to form $N$-particle charge correlations, may be guided by the distributions of events in the relative momentum space. The study of leading dihadron correlations with respect to the full kinematic distribution \cite{Dreyer:2018nbf} and the relation to perturbative jet showers through jet declustering and grooming \cite{Larkoski:2014wba, Dreyer:2018tjj, Frye:2017yrw} is ongoing. Promoting our hadronic studies to subjet charge correlation among leading subjets is also a promising direction.  

More generally, understanding the flow of flavor within jets will require high-precision observations in momentum space, like those we have discussed above, supplemented by strong capabilities in particle identification.  We believe that this is a promising approach toward a deeper understanding of the transition from partonic to hadronic degrees of freedom in Quantum Chromodynamics.

\section{Acknowledgements}

The authors thank Miguel Arratia, Henry Klest, Raghav Kunnawalkam Elayavalli, Petar Maksimovic, Simone Marzani, Brian Page, Felix Ringer, Gregory Soyez for useful comments and discussions. Y.-T.C. would like to thank Yu-Chen Chen for the help with Herwig simulation which is performed on the Grendel cluster at Massachusetts Institute of Technology. This work of Y.-T.C.\ and G.S.\ was supported in part by the National Science Foundation, award PHY-1915093.

\bibliography{ref}
\end{document}